\documentclass[11pt,letterpaper]{article}
\usepackage[utf8]{inputenc}
\usepackage[T1]{fontenc}
\usepackage{poma_style}
\usepackage{graphicx}
\usepackage{amsmath,amsfonts,bm}
\usepackage{booktabs}
\usepackage{siunitx}
\usepackage{microtype}
\usepackage{xcolor}
\usepackage{array}

\usepackage[colorlinks=true,urlcolor=blue,linkcolor=black,citecolor=black]{hyperref}
\hypersetup{pageanchor=false}
\newcommand{\sphere}{\mathbb{S}^{2}}
\newcommand{\jsdist}{d_{\mathrm{JS}}}
\newcommand{\kld}{D_{\mathrm{KL}}}

\title{}
\author{}
\date{}

\begin{document}

\begin{center}
{\LARGE\bfseries
A Compact Statistical View of
Musical-Instrument Directivity\par}

\vspace{1em}

{\large
Gioele Greco, Mirco Pezzoli,
Raffaele Malvermi,
Gian Marco Ricci, and Fabio Antonacci\par \normalsize Politecnico di Milano, Milano, Italy\\
\normalsize \texttt{gioele.greco@polimi.it}}

\vspace{0.8em}


\vspace{0.5em}

{\small
International Symposium on Musical Acoustics
(ISMA 2026)\\
Helsinki, Finland, 15--17 June 2026
\par}
\end{center}

\vspace{1em}

\section*{Abstract}
Directivity is a defining feature of musical-instrument radiation and is known to vary strongly with frequency. With the growing availability of measured directivity datasets and their use in sound-field simulation and spatial audio, there is a practical need for representations that are both compact and comparable across sources. This work addresses that need by proposing a statistically grounded description of frequency-dependent directivity that reduces complex radiation patterns to a small set of meaningful descriptors. The approach enables directivity patterns to be summarised in an interpretable way and compared consistently across frequency bands, even when the radiation is not well described by a single dominant direction. The resulting representation supports tasks such as analysis, categorisation, and model building, providing a bridge between detailed measurements and higher-level descriptors useful for musical acoustics and audio rendering. Experiments on a corpus of 41 orchestral instruments show that the proposed method preserves the distance geometry well while reducing the full order-4, 30-band corpus from about 492~kB to about 100~kB in double-precision storage.

\noindent\textbf{Author Keywords:} musical instrument directivity; spherical harmonics; directional statistics; von Mises-Fisher mixture; clustering.

\section{Introduction}
The directional radiation of a musical instrument shapes how spectral energy reaches the room, how timbre varies with listening position, and how instruments interact inside an ensemble and in a performance space. Directivity therefore belongs to the source itself: it expresses a frequency-dependent aspect of the instrument's radiating mechanism and contributes to projection, blend, and spatial impression \cite{Meyer1972,Patynen2010,Corcuera2023,Ackermann2024dynamic}. Contemporary directivity datasets make this structure available at corpus scale, with frequency-resolved measurements for complete families of orchestral instruments \cite{Ackermann2024}. That availability shifts the central question from measurement alone to representation.
Full spherical maps remain the physically richest representation. They are also expensive objects. They occupy many coefficients or many sampled directions per band, they are awkward to browse across large corpora, and they make repeated comparison costly. This becomes a practical bottleneck as soon as directivity is used beyond visual inspection. Corpus studies, similarity search, retrieval, surrogate modeling, and interactive spatial-audio workflows all benefit from descriptors that retain acoustically meaningful structure while remaining compact enough for repeated statistical operations.

This requirement aligns with a broader tendency in audio and music research. Many successful machine-learning and retrieval systems rely on compressed, stable, and interpretable feature spaces rather than on direct operation over raw high-dimensional measurements \cite{Muller2011,Cobos2022,Chen2024}. The same logic applies to directional data. A useful directivity descriptor should preserve the organization of radiation across frequency, support comparison across instruments, and scale naturally to dataset-level analysis.

Existing work has addressed important parts of the problem. Robust reconstruction methods have improved the estimation of radiation patterns from sparse spherical measurements \cite{Canclini2020}. Pattern-level similarity measures, including spherical correlation and retrieval-oriented distances, have made inter-instrument comparison more systematic \cite{Carpentier2023,Pezzoli2023,Pezzoli2022}. In parallel, directional-statistical models based on the von Mises-Fisher (vMF) distribution have proved useful in other acoustical problems, including robust directional beamforming \cite{Anderson2015} and the analysis of reverberant fields captured with spherical microphone arrays \cite{Bastine2020,Bastine2022}. In particular, the recent works \cite{Bastine2020,Bastine2022} show that vMF mixtures can summarize directional power distributions estimated from spherical-array measurements in a compact and physically meaningful way. Those studies, however, target reverberation and reflection analysis rather than source directivity as an object to be compressed and compared across a corpus. A compact statistical representation therefore remains valuable because it turns a dense source-radiation map into a small set of acoustically readable parameters.

The present paper develops that representation. Each band-limited directivity pattern is interpreted as a directional energy distribution on the sphere and approximated by a small mixture of vMF components. Each retained component carries a clear physical meaning: a dominant radiation direction, a concentration that reflects lobe sharpness, and a weight that quantifies relative prominence. The result is a compact, frequency-aware description that supports compression and comparison at the same time. The paper focuses on three questions: how much the corpus can be compressed, how accurately pairwise relations are preserved after compression, and how the resulting descriptors organize instruments in a reduced feature space.

\section{Method}
\subsection{Directivity signal model}
\begin{figure}[t]
    \centering
    \includegraphics[width=0.75\linewidth]{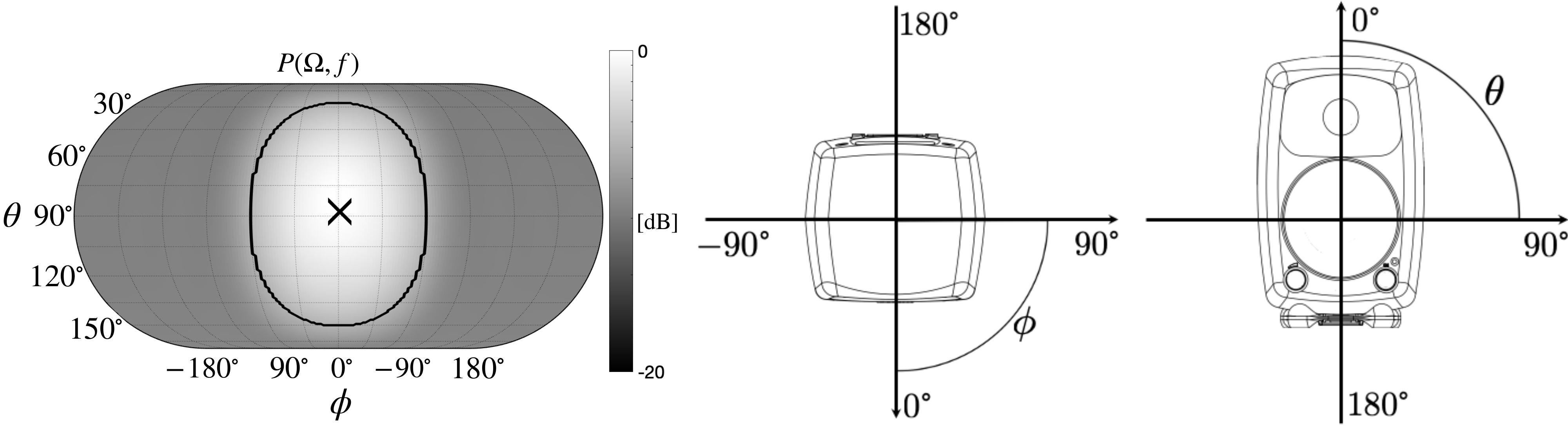}
    \caption{Example of a directivity pattern of a Genelec 8030A loudspeaker at $\SI{1.4}{\kilo\hertz}$ taken from \cite{tylka2015database}. The black line indicates the \SI{-3}{\decibel} threshold, while the cross shows the center of mass. The reference system is centered on the loudspeaker woofer.}
    \label{fig:directional}
\end{figure}

For an acoustic source in free-field conditions, the directional dependence of sound radiation can be expressed through the complex directivity function $D(\Omega,f)$ \cite{Ackermann2024dynamic}, where $\Omega=(\phi,\theta)$ denotes direction on the sphere and $f$ denotes frequency, an example is depicted Figure~\ref{fig:directional}. 
In musical-acoustics practice, the terms \emph{directivity}, \emph{directivity pattern}, and \emph{radiation pattern} are sometimes used with slightly different conventions: depending on context they may refer to the complex transfer function itself, to its magnitude, or to a level/power map derived from it \cite{Pezzoli2022,Pezzoli2023,Ackermann2024}. In the present paper the goal is statistical modeling of directional energy, so we work with a nonnegative bandwise quantity that can be normalized into a distribution over $\sphere$.
Using SH analysis order $L$, the directivity function and the bandwise directional energy pattern are written as \cite{Canclini2020}
\begin{equation}
    D(\Omega,f)=\sum_{\ell=0}^{L}\sum_{m=-\ell}^{\ell} c_{\ell m}(f)Y_{\ell m}(\Omega),
\label{eq:sh-expansion}
\end{equation}
where $Y_{\ell m}$ is the spherical harmonics of order $\ell$ and degree $m$ \cite{schmitz2012using}, and $C_{\ell m}(f)$ is the corresponding coefficient. 
In this work, we are interested in modeling the so-called directivity pattern, that is the bandwise directional energy defined as
\begin{equation}
    P(\Omega,f)=\left|D(\Omega,f)\right|^2.
\label{eq:dirctivity_pattern}
\end{equation}

This choice makes the modeled quantity directly interpretable as angular energy density and matches the notion of a directivity pattern as a directional sound-radiation map \cite{Pezzoli2022,Pezzoli2023}. In practice, $P(\Omega,f)$ is sampled on a spherical grid $\{\bm{x}_n\}_{n=1}^{N}\subset\sphere$ and normalized to unit mass as 
\begin{equation}
p_n(f)=\frac{P_n(f)}{\sum_{m=1}^{N}P_m(f)},
\qquad P_n(f)=P(\Omega_n,f),
\label{eq:pdf}
\end{equation}
so that overall level is removed and only the angular distribution of radiated energy is retained.

\subsection{Direcitivity as von Mises-FIsher mixtures}
The use of vMF models in this work is motivated both by directional statistics and by prior acoustical applications. Beyond their general role as canonical distributions on the sphere \cite{MardiaJupp2000}, vMF mixtures have been used to cluster or parameterize directional data on hyperspheres \cite{Banerjee2005} and to describe directional sound fields derived from spherical-array measurements \cite{Anderson2015,Bastine2020,Bastine2022} by a small set of parameters.
Therefore, they represent a compact representation for directivity \eqref{eq:dirctivity_pattern} in terms. 
Let us consider a single vMF density on $\sphere$ defined as
\begin{equation}
f(\bm{x};\bm{\mu},\kappa)=c_3(\kappa)\exp\!\left(\kappa\,\bm{\mu}^{\mathsf T}\bm{x}\right),
\qquad \bm{x}\in\sphere,
\label{eq:single-vmf}
\end{equation}
where $\bm{\mu}\in\sphere$ is the mean direction and $\kappa\ge 0$ is the concentration parameter \cite{Fisher1953,MardiaJupp2000,Banerjee2005}. 
In three dimensions the normalization constant is
\begin{equation}
c_3(\kappa)=\frac{\kappa}{4\pi\sinh\kappa},
\label{eq:c3}
\end{equation}
with isotropic limit $c_3(0)=1/(4\pi)$. The density is rotationally symmetric around $\bm{\mu}$: when $\kappa=0$ it becomes uniform on the sphere, while increasing $\kappa$ produces progressively sharper directional concentration. 
Thus, considering a high-energy radiation sector in $p_n(f)$, the main direction of emission is indicated by $\bm\mu$, while the size of the sector is represented by $\kappa$. 
It follows, that the frequency-dependent directivity pattern \eqref{eq:pdf} can be represented as a weighted sum of $K$ vMF densities \eqref{eq:single-vmf} as
\begin{equation}
\hat p_K(\bm{x}\,|\,f)=\sum_{k=1}^{K}\pi_k(f)\,c_3\!\left(\kappa_k(f)\right)
\exp\!\left(\kappa_k(f)\,\bm{\mu}_k^{\mathsf T}(f)\bm{x}\right),
\label{eq:vmf-mixture}
\end{equation}
where $\pi_k\ge 0$, $\sum_k \pi_k=1$, $\bm{\mu}_k\in\sphere$, and $\kappa_k\ge 0$ \cite{Banerjee2005}. In this representation, each component models one radiating lobe: $\bm{\mu}_k$ identifies its main direction, $\kappa_k$ describes its angular sharpness, and $\pi_k$ quantifies the relative share of radiated energy associated with that lobe. Mixtures are therefore a natural fit when the radiation is multimodal and cannot be summarized by a single dominant direction.


For a fixed $K$, the parameters are estimated by weighted expectation-maximization from the sampled spherical distribution. The selected order is the smallest $K$ for which the reconstructed distribution reaches a prescribed spherical Pearson similarity with the raw one. With $\bm{p}(f)$ and $\hat{\bm{p}}_K(f)$ denoting the sampled raw and reconstructed distributions,
\begin{equation}
\rho\!\left(\bm{p},\hat{\bm{p}}\right)=
\frac{(\bm{p}-\bar p\bm{1})^{\mathsf T}(\hat{\bm{p}}-\bar{\hat p}\bm{1})}
{\lVert \bm{p}-\bar p\bm{1}\rVert_2\,\lVert \hat{\bm{p}}-\bar{\hat p}\bm{1}\rVert_2},
\label{eq:sph-pearson}
\end{equation}
and the retained order is
\begin{equation}
K^{(\tau)}(f)=\min\!\left\{K:\rho\!\left(\bm{p}(f),\hat{\bm{p}}_K(f)\right)\ge \tau\right\}.
\label{eq:kselect}
\end{equation}
The target similarity $\tau$ therefore acts as a direct control of representation fidelity.

\subsection{Compression accounting}
One of the main advantages of the vMF mixture representation is represented by its compact model. 
Let us consider the size of the SH representation of one valid frequency band at order $L$. 
Being complex numbers, SH coefficients are stored as $2(L+1)^2$ real scalars. It follows that for an instrument with $B_i$ bands the total number of real values is
\begin{equation}
R_i=2(L+1)^2B_i.
\label{eq:rawcount}
\end{equation}
The compressed vMF representation stores one validity flag per active band and, for each retained vMF component, one weight, one concentration, and three Cartesian coordinates for its mean direction,
\begin{equation}
C_i^{(\tau)}=B_i+5\sum_{f\in\mathcal{B}_i}K_i^{(\tau)}(f).
\label{eq:compressedcount}
\end{equation}
The compression factor is defined as $\Gamma_i^{(\tau)}=\frac{R_i}{C_i^{(\tau)}}.$

\section{Experimental protocol}
The experiments use the public database of directivities introduced by Ackermann, Brinkmann, and Weinzierl \cite{Ackermann2024}. The corpus contains 41 modern and historical orchestral sources measured in the anechoic chamber of Technische Universit\"at Berlin with a 32-channel full-spherical microphone array. The released material includes frequency-averaged directivities in one-third-octave bands from \SI{25}{Hz} to \SI{20}{kHz}, stored in SOFA format at \SI{44.1}{kHz}, together with single-tone measurements and metadata on instrument identity and pitch range \cite{Ackermann2024}. The room volume is approximately \SI{1070}{m^3} and the low-frequency chamber cutoff is around \SI{63}{Hz} \cite{Ackermann2024}. These characteristics make the dataset suitable for musical-acoustics analysis and for corpus-level evaluation of compact directivity descriptors.

The SH analysis order is fixed to $L=4$, which matches the highest order commonly supported by a 32-channel spherical array under standard sampling assumptions. At this order, each valid band contains 25 complex SH coefficients, that is 50 real scalars. For the full 41-instrument, 30-band corpus, the raw SH description therefore contains $41\times 30\times 50=61{,}500$ real values. Under double-precision storage this corresponds to about 492~kB. This byte count refers to the analytical SH representation used in the paper rather than to the original SOFA containers. The mean compression factors reported in Figure~\ref{fig:compression} translate this corpus-level size to about 78~kB at 50\%, 100~kB at 70\%, 197~kB at 90\%, and 289~kB at 95\%. The detailed comparative analysis uses the subset of bands represented by at least three instruments, which stabilizes corpus averages at the spectral extremes.

Each valid band is resampled on a common 720-point Fibonacci sphere and converted into the normalized directional energy distribution of \eqref{eq:pdf}. Bandwise vMF fitting is carried out independently across frequency. This choice matches the acoustical behavior of musical instruments: wavelength, bore geometry, tone holes, bell radiation, string-body coupling, and structural vibration all reshape the directional field with frequency, so each band benefits from its own local model. Weighted expectation-maximization is run with a maximum of 200 iterations, a relative log-likelihood tolerance of $10^{-7}$, concentration bounds $[10^{-4},120]$, and a maximum of 20 components per band. Candidate mean directions are initialized from directional peaks with minimum angular separation constraints to encourage distinct lobes.

Four target fidelities are evaluated, $\tau\in\{0.50,0.70,0.90,0.95\}$. The 70\% condition receives the most attention because it provides the clearest balance between compactness and structural preservation in the present corpus. Inter-instrument distances are computed bandwise from the normalized directional energy distributions and then aggregated across overlapping valid bands with uniform band weights and a coverage penalty for incomplete overlap. The explicit Jensen-Shannon definition used for this comparison is introduced in Sec.~\ref{sec:js_results}. Hierarchical clustering uses average linkage. For low-dimensional inspection, the fixed-length permutation-invariant descriptors introduced in Sec.~\ref{sec:descriptor_summaries} are standardized and projected by principal-component analysis.
Table~\ref{tab:abbr} expands the abbreviated labels that appear in the figures. The table supports direct reading of the dendrograms, distance matrices, and principal-component plot.

\begin{table}[t]
\centering
\caption{Abbreviated labels used in the figures.}
\label{tab:abbr}
\scalebox{0.90}{
\scriptsize
\setlength{\tabcolsep}{3pt}
\begin{tabular}{ll@{\hspace{0.8em}}ll@{\hspace{0.8em}}ll}
\toprule
Abbr. & Instrument & Abbr. & Instrument & Abbr. & Instrument \\
\midrule
A.SaxM & alto saxophone (modern) & ClM & clarinet (modern) & ObM & oboe (modern) \\
A.TbnH & alto trombone (historical) & DouM & double-action harp (modern) & S.Sop & soprano \\
AcoM & acoustic guitar (modern) & DulH & dulcian (historical) & T.SaxM & tenor saxophone (modern) \\
B.BsnH & classic bassoon (historical) & FlH & transverse flute (historical) & T.TbnH & tenor trombone (historical) \\
B.BsnH2 & baroque bassoon (historical) & FlH2 & keyed flute (historical) & T.TbnM & tenor trombone (modern) \\
B.BsnM & bassoon (modern) & FlM & flute (modern) & TbaM & tuba (modern) \\
B.ClM & bass clarinet (modern) & HnH & natural horn (historical) & TptH & natural trumpet (historical) \\
B.DouH & double bass (historical) & HnM & French horn (modern) & TptM & trumpet (modern) \\
B.DouM & double bass (modern) & HnM2 & English horn (modern) & VcH & cello (historical) \\
B.HnH & basset horn (historical) & ObH & classic oboe (historical) & VcM & cello (modern) \\
B.TbnH & bass trombone (historical) & ObH2 & romantic oboe (historical) & VlaH & viola (historical) \\
B.TbnM & bass trombone (modern) & ClH & clarinet (historical) & VlaM & viola (modern) \\
Cb.BsnM & contrabassoon (modern) & VlnH & violin (historical) & VlnM & violin (modern) \\
\bottomrule
\end{tabular}
}
\end{table}

\section{Results}
\subsection{Compression and model complexity across frequency}
\begin{figure}[t]
    \centering
    \includegraphics[width=0.9\linewidth]{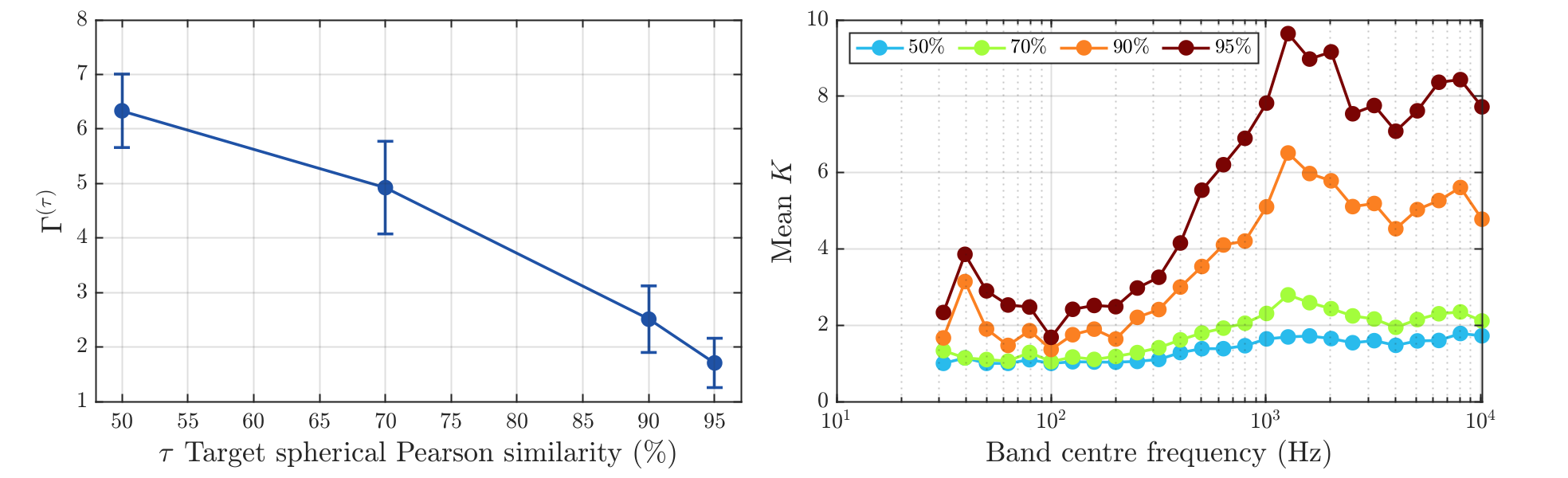}
    \caption{Compression-fidelity trade-off. Left: mean compression factor, defined as raw scalars divided by compressed scalars, with standard-deviation error bars. Right: mean minimum number of vMF components required per one-third-octave band. Component demand rises mainly in the upper bands and especially at the stricter fidelity targets.}
    \label{fig:compression}
\end{figure}

Figure~\ref{fig:compression} states the core result of the paper. The mean compression factor is about 6.3 at 50\%, 4.9 at 70\%, 2.5 at 90\%, and 1.7 at 95\%. The 70\% operating point therefore reduces the full order-4 corpus from about 492~kB to about 100~kB while keeping an explicit directional model in every active band. At that operating point, one band is commonly represented with two or three vMF components instead of 25 complex SH coefficients.

The right panel of Figure~\ref{fig:compression} explains where the degrees of freedom are spent. Below a few hundred hertz, the mean minimum number of components stays close to one for the relaxed and intermediate targets, which reflects the broad radiation expected when wavelength is large with respect to the main radiating dimensions. Complexity grows gradually through the mid band and rises sharply above about \SI{1}{kHz}, where secondary lobes and family-specific angular detail become stronger. The 95\% condition reaches roughly eight to ten components in the upper bands, whereas the 70\% condition remains around two to three components. This is the practical signature of the trade-off: most of the additional budget beyond 70\% is devoted to upper-band fine structure.

\subsection{Preservation of pairwise relations}
\label{sec:js_results}
To compare instruments, the Jensen-Shannon distance is evaluated on the \emph{bandwise normalized directional energy distributions} defined in \eqref{eq:pdf}. More precisely, for instrument $i$ and band $f$, the raw object is the discrete spherical distribution $\bm{p}_i(f)=\bigl[p_{i,1}(f),\dots,p_{i,N}(f)\bigr]^{\mathsf T}$ obtained from the SH-based energy pattern $P_i(\Omega,f)=|D_i(\Omega,f)|^2$ after sampling and normalization. After vMF fitting, the corresponding reconstructed object is $\hat{\bm{p}}_i(f)=\bigl[\hat p_i(\bm{x}_1|f),\dots,\hat p_i(\bm{x}_N|f)\bigr]^{\mathsf T}$, evaluated on the same spherical grid. For any two same-band distributions $\bm{a}$ and $\bm{b}$, either both raw or both reconstructed, we use
\begin{equation}
\jsdist(\bm{a},\bm{b})=
\sqrt{\frac{1}{2}\kld(\bm{a}\,\|\,\bm{m})+\frac{1}{2}\kld(\bm{b}\,\|\,\bm{m})},
\qquad
\bm{m}=\frac{1}{2}(\bm{a}+\bm{b}),
\label{eq:jsd}
\end{equation}
with $\kld(\bm{a}\,\|\,\bm{m})=\sum_n a_n\log\!\bigl(a_n/m_n\bigr)$ and the convention $0\log 0=0$ \cite{Lin1991}. The Jensen-Shannon divergence is symmetric, bounded, and always finite; its square root is a true metric \cite{Endres2003,Fuglede2004}.

Specifically, for instrument pair $(i,j)$ and band $f$, we compare either the raw distributions
$\bm a=\bm p_i(f)$ and $\bm b=\bm p_j(f)$,
or the reconstructed distributions
$\bm a=\hat{\bm p}_i(f)$ and $\bm b=\hat{\bm p}_j(f)$,
all evaluated on the same spherical grid. We adopt it here because the compared objects are probability-like directional energy distributions on the sphere, not generic signed vectors. In other words, the metric directly measures how differently two instruments distribute radiated energy over angle. This is more appropriate than normalized cross correlation (NCC), which measures centered linear agreement and has been mainly used as a similarity score for pattern matching in coefficient or spatial domains \cite{Carpentier2023,Pezzoli2022,Pezzoli2023}.

\begin{figure}[h]
    \centering   \includegraphics[width=1\linewidth]{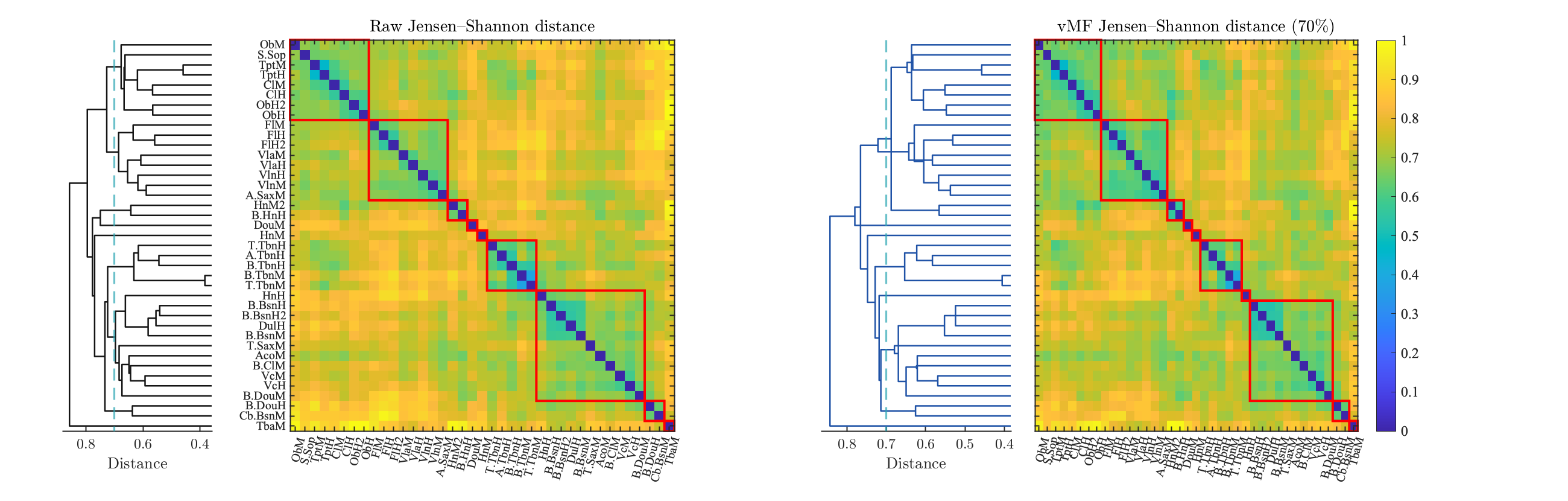}
    \caption{Pairwise distance geometry at the 70\% operating point. Left: raw Jensen-Shannon distance matrix. Right: distance matrix obtained from the reconstructed vMF representation.}
    \label{fig:distances}
\end{figure}
Fig.~\ref{fig:distances} show how much of the original corpus geometry survives after compression. The 70\% dendrogram retains the large-scale grouping already visible in the raw distance structure. Neighbor relations and branch lengths change locally, yet the broad organization remains stable enough to support family- and register-level interpretation.
Quantitatively, the agreement between the upper-triangular matrix entries reaches Pearson 0.961, with RMSE 0.093. These values support the use of the compressed representation for retrieval, clustering, and dataset navigation.

\subsection{Permutation-invariant descriptor construction}
\label{sec:descriptor_summaries}

While the fitted mixtures represent the radiation patterns compactly, expectation-maximization (EM) assigns arbitrary indices to the mixture components. Because this ordering varies across frequency bands and instruments, direct comparison of the raw parameters is not meaningful \cite{Stephens2000}. To enable consistent corpus-level visualization (e.g., via PCA), we convert each bandwise mixture into a fixed-length set of permutation-invariant descriptors.

For each band, we compute the effective number of components, $N_{\mathrm{eff}}=\exp\!\left(-\sum_k \pi_k\log\pi_k\right)$; the resultant vector, $\bm{r}=\sum_k \pi_k A_3(\kappa_k)\bm{\mu}_k$, where $A_3(\kappa) = \coth(\kappa) - 1/\kappa$ maps the concentration $\kappa$ to the mean resultant length; and the second-order directional moment tensor,
\begin{equation}
\bm{M}=\sum_k \pi_k\bm{\mu}_k\bm{\mu}_k^{\mathsf T}.
\label{eq:moment}
\end{equation}
Together, these quantities capture complementary aspects of the radiation, lobe complexity ($N_{\mathrm{eff}}$), overall directional bias ($\bm{r}$), and spatial spread or anisotropy ($\bm{M}$), without depending on component order. After concatenation across frequencies, these fixed-length summaries provide a canonical representation used in this work to inspect corpus- and family-level structures.

\begin{figure}[t]
    \centering
    \includegraphics[width=0.51\linewidth]{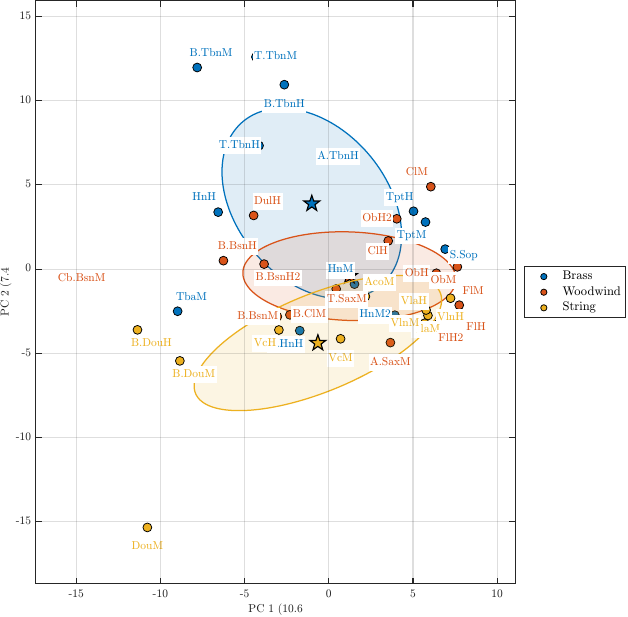}
    \caption{Principal-component projection of the standardized invariant descriptors at the 70\% operating point. Colors indicate broad instrument families, translucent convex hulls summarize their occupied regions, and stars denote family centroids.}
    \label{fig:pca}
\end{figure}

Figure~\ref{fig:pca} summarizes the permutation-invariant descriptors at the 70\% operating point. The first two principal components explain 10.6\% and 7.4\% of the standardized variance. This projection is used as a reading aid for the descriptor space. Brass instruments occupy a higher region of the second principal component and span a wider vertical extent, woodwinds concentrate more tightly around the center, and strings populate a partially overlapping lower-right region. The overlap between family hulls is informative: directivity reflects organological family together with register, construction, and radiating mechanism. The compact descriptors retain both the broad family trend and the individuality of specific instruments.

\section{Discussion}
The results identify a useful operating region for compact directivity modeling. Around the 70\% target similarity, the representation stores about one fifth of the raw SH corpus, preserves the inter-instrument distance geometry with high correlation, and keeps a direct physical interpretation through lobe directions, concentrations, and weights. The stricter targets allocate most of the additional parameters to upper-band detail. This makes them attractive for near-literal reconstruction and less attractive for corpus analysis, where relational structure carries the main information.
The representation also provides a stable interface to data-driven workflows. Its parameters remain interpretable, its descriptors are permutation invariant, and its storage cost is low enough for repeated comparison across a corpus. These properties are useful for musical-acoustics studies, for directivity-aware retrieval and clustering, and for future learning systems that require compact directional features. The method therefore serves both as a compression strategy and as a descriptive layer for frequency-dependent source radiation.

\section{Conclusion}
A compact statistical representation of musical-instrument directivity has been presented based on bandwise mixtures of von Mises-Fisher distributions fitted to spherical energy distributions reconstructed from SH data. The model makes complexity directly controllable through a target similarity and expresses each band with a small set of directional descriptors.
On the analyzed orchestral corpus, the 70\% operating point provides the clearest balance between size and fidelity. It reduces the full order-4, 30-band corpus from about 492~kB to about 100~kB in double precision and preserves the Jensen-Shannon distance geometry with high correlation. These results show that a limited set of directional-statistical parameters captures a large share of the structure that matters for comparing musical-instrument radiation.
Future work can extend the model with continuity constraints across frequency, integrate playing conditions and note dependence more explicitly, and use the resulting descriptors in retrieval and learning-oriented directivity models for musical acoustics and spatial audio.

\bibliographystyle{unsrt}
\bibliography{refs_merged}

\end{document}